\documentclass[a4paper]{revtex4}
\usepackage{amsmath,amssymb,amsthm,amscd}
\usepackage{graphicx}
\usepackage[colorlinks]{hyperref}

\begin{document}

\title {Production of primordial gravitational waves in a simple class of running vacuum cosmologies}

\author{D. A. Tamayo}\email{tamayo@if.usp.br}
\affiliation{Instituto de F\'{i}sica, Universidade de S\~ao Paulo, Rua do Mat\~ao 187, CEP 05508-090, S\~ao Paulo, SP, Brazil}

\author{J. A. S. Lima,}\email{limajas@astro.iag.usp.br}
\affiliation{Departamento de Astronomia, Universidade de S\~ao Paulo, Rua do Mat\~ao 1226, 05508-900, S\~ao Paulo, Brazil}

\author{D. F. A. Bessada}\email{dennis.bessada@unifesp.br}
\affiliation{Laborat\'orio de F\'\i sica Te\'orica e Computa\c c\~ao Cient\'\i fica, Universidade Federal de S\~ao Paulo - UNIFESP, Campus Diadema, Brazil}

\begin{abstract}
The problem of cosmological production of gravitational waves  is discussed in the framework of an expanding, spatially homogeneous and isotropic FRW type Universe with time-evolving vacuum energy density. The gravitational wave equation is established and its modified time-dependent part is analytically resolved for different epochs in the case of a flat geometry. Unlike the standard $\Lambda$CDM cosmology (no interacting vacuum), we show that  gravitational waves are produced in the radiation era even in the context of general relativity. We also show that for all values of the free parameter, the high frequency modes are damped out even faster than in the standard cosmology  both in the radiation and matter-vacuum dominated epoch. The formation of the stochastic background of gravitons and the remnant power spectrum generated at different cosmological eras are also explicitly evaluated. It is argued that measurements of the CMB polarization (B-modes) and its comparison  with the rigid $\Lambda$CDM model plus the inflationary paradigm may become a crucial test for dynamical dark energy models in the near future.\\

\end{abstract}
\maketitle

\section{Introduction}

The recent direct detection of gravitational waves (GWs) by the LIGO collaboration \cite{LIGO2016} is a turning point in the field opening the new era of GW astronomy. This result provoked both surprise and excitement not only on the researchers working with the generation of GWs from astronomical objects but also of cosmological origin. The implications are promising because the weak interaction of the GWs with matter will provide information previously unattainable like the physics of strong relativistic gravitational fields (black holes coalescence), specific signatures from the very early universe (much before CMB), and even tests of gravitational theories\cite{LIGOGRtest}. Potentially, cosmological GWs will furnish new and powerful key for accessing the early universe physics and their current problems like the mystery related with the nature of the unknown components in the Universe (dark matter and dark energy).

In this concern, the simplest explanation for the present accelerating stage of the observed Universe is the existence of a dark energy (DE) component,  in addition to the cold dark matter (CDM) \cite{Copeland2006,Frieman2008,Bartelmann2010}. Observationally, the most accepted candidate for DE is a rigid cosmological constant          $\Lambda$ with energy density $\rho_{\text{vac}} = \Lambda/8\pi G$. At the level of general relativity (GR) and quantum field theory (QFT) in curved spacetimes, the $\Lambda$-term can be interpreted as a decoupled relativistic simple fluid (vacuum medium) obeying the equation of state (EoS), $p_{\text{vac}} = - \rho_{\text{vac}}=constant$ \cite{Zeldovich1967}.

The recent Planck2015 results confirmed that the cosmic concordance model ($\Lambda$CDM) - a flat Universe filled with cold dark matter (CDM), baryons and a $\Lambda$-term - is in good agreement with the currently available cosmological observations \cite{Ade2015}. However, from the theoretical viewpoint,  the unsettled situation in the particle physics/cosmology interface in which the cosmological upper bound ($\rho_{\text{vac}} \lesssim 10^{-47} GeV^4$) differs from naive theoretical QFT expectations ($\rho_{\text{vac}} \sim 10^{71} GeV^4$) by more than 100 orders of magnitude, originates an extreme fine-tuning problem, the so called cosmological constant problem \cite{Zeldovich1967, Weinberg1989}. Another unsolved mystery from first principles is why the  vacuum energy density is so close to the matter energy density which is sometimes referred to as cosmic coincidence problem \cite{Peebles2003,Padmanabhan2003,Lima2004}. 

In order to alleviate the cosmological constant and coincidence problems, and also some mild tensions of the $\Lambda$CDM model \cite{Amendola10,LCunha14,Grandis16} many authors have proposed dynamical DE models (for a recent discussion see \cite{Sahni2014}). In general grounds, these models can also be implemented  trough a time-evolving vacuum energy density coupled with the remaining cosmic components. Theoretically, such models are suggested by the general form of the effective action in QFT in curved spacetimes \cite{Shapiro2002}.  However, there are also some attempts to represent such $\Lambda$ models by a scalar field \cite{Maia2002,Bessada2013}, a Lagrangian description  based on the so-called $F(R,T)$ gravity where $R$ is the curvature scalar and $T$ is the trace of the energy-momentum tensor \cite{Poplawski2006,Harko}.  In such frameworks, different phenomenological decay laws for a time-dependent $\Lambda$ have also been proposed and their predictions confronted with the available observational data \cite{Ozer1986, Carvalho1992, Waga1993, Lima1994, Lima1996, Overduin1998, Carneiro2005, Lima2013, Perico2013, Basilakos2012, Sola2016, LBS2016}. 

Broadly speaking, the basic idea of a time-evolving $\Lambda$-term is quite simple. The vacuum EoS implies that $\rho_{\text{vac}}$ is constant only when the vacuum component is separately conserved. This is not the case when an interacting mixture is considered as in the present work. To be more specific, consider the corresponding Einstein field equations for a system formed by the interacting mixture (matter plus the vacuum fluid) with energy momentum tensors  $T^{\mu\nu}_{(\text{mat})}$ and $T^{\mu\nu}_{(\text{vac})}$, respectively. Since the Einstein tensor $G^{\mu \nu}$ is divergenceless, the total energy conservation law in the direction of the four-velocity, $u_{\mu}(T^{\mu\nu}_{(\text{mat})} + T^{\mu\nu}_{(\text{vac})})_{;\nu}=0$, reads:

\begin{equation}\label{ECL}
u_{\mu}T^{\mu\nu}_{(\text{mat}){;\nu}} = {\dot\rho} + 3H (\rho + p) = -{\dot\rho}_{\text{vac}}  \equiv - \frac{\dot \Lambda(t)}{8\pi G}\,,
\end{equation}
where a FRW geometry was assumed and in the last equality the vacuum EoS was used (see (\ref{eq4})). Note that the energy of the material component  is not conserved independently. The running vacuum component is continuously transferring energy to the fluid component. Of course, when the energy density of the vacuum medium  becomes constant it does not contribute to the above balance equation for the material component. In other words, if $\Lambda(t)$ is the constant (bare) $\Lambda_b$, the energy of the material medium is separately conserved thereby recovering the standard result.

On the other hand, the recent LIGO results are pressing the search on the  production of  relic GWs (tensor modes) both at very low and high frequencies. In the observational front,  the major challenge is the detection of the B-modes polarization anisotropies in the cosmic background radiation (CMB) provoked by tensor fluctuations produced in the inflationary regime. The expected level of such anisotropies is small but presumably it will be accessed by the ongoing (BICEP and Keck Array experiments, South Pole Telescope, etc.) and future probes through a new generation of high sensitivity instruments. One of them is the QU bolometric interferometer for cosmology  (QUIBIC) whose basic goal is to measure the B-modes at angular intermediate scales based on the innovative technological concept of bolometric interferometry (see, for instance, \cite{QUBIC}). In principle, the presence of a time-evolving $\Lambda$-term must effect the B-modes in comparison with the traditional $\Lambda$CDM plus the inflationary paradigm.  As a consequence, the influence of a dynamical vacuum on the adiabatic amplification of GWs is a possibility that needs to be investigated. This is one of the main aims of the present paper. Recent similar studies have appear in the literature Similar recent studies have appeared in the literature \cite{Tamayo}.   

In this context we investigate here the production of primordial GWs in the framework of a simple $\Lambda(H)$ decaying vacuum cosmology that recovers the $\Lambda$CDM model as a particular case. The energy density and power spectrum are analytically derived. Interestingly, unlike the cosmic concordance model ($\Lambda$CDM), there exists adiabatic amplification of GWs during the radiation epoch. As we shall see, such a result holds for generic decaying vacuum models and must remain valid for all analyzes based on realistic running vacuum cosmologies.

\section{The Model: Basic Equations}

Let us now consider that the Universe is well described by a flat FRW geometry. In the co-moving coordinate
system, the background line element reads ($c=1$):

	\begin{equation} \label{eq1}
	ds^2 = -dt^2 + a^2(t)\,dl^2 = a^2(\eta)\,(-d\eta^2 + dl^2)\;,
	\end{equation}
where the 3-space metric is $dl^{2}=dx^{2} + dy^{2} + dz^{2}\equiv\delta_{ij}dx^{i}dx^{j}$, $i,j,k=1,2,3$ and $t$, $\eta$ are, respectively, the cosmic and conformal times, related by

	\begin{equation} \label{eq2}
	dt=a\,d\eta\;.
	\end{equation}
The Einstein field equations for a  two-fluid mixture (matter + vacuum) is given by:

	\begin{equation} \label{eq3}
	R^{\mu\nu}-\frac{1}{2}\,R\,g^{\mu\nu} =8\pi G(\,T^{\mu\nu}_{\text{mat}}+ T^{\mu\nu}_{\text{vac}})\;,
	\end{equation}
with $R_{\mu\nu}$ and $R$, being the Ricci tensor and Ricci scalar, respectively. The energy-momentum tensors (matter and vacuum) read:

	\begin{equation} \label{eq4}
	T^{\mu\nu}_{\text{mat}} = (\rho+p)\,u^{\mu}\,u^{\nu}-p\,g^{\mu\nu},\quad T^{\mu\nu}_{\text{vac}}=\rho_{\text{vac}}\,g^{\mu\nu}\,,
	\end{equation}
where $\rho$, $p$, $u^{\mu}$ are, respectively, the energy density, pressure and four-velocity of the fluid, whereas $\rho_{\text{vac}} \equiv \Lambda(t)/8\pi G$, is the vacuum energy density. The field equations in the above background take the form \cite{Carvalho1992, Lima1994, Overduin1998}

	\begin{eqnarray}
	8\pi G\,\rho\,+\,\Lambda(t) &=& 3\frac{\dot{a}^2}{a^2}\;,\label{eq5}\\
	8\pi\, G\, p - \Lambda(t) &=& -2\,\frac{\ddot{a}}{a}\,-\,\frac{\dot{a}^2}{a^2}\;,\label{eq6}
	\end{eqnarray}
where a dot means derivative with respect to the cosmic time $t$. In order to solve the above equation we need to know both the functional form of $\Lambda(t)$ and the EoS.

Many phenomenological functional forms have been proposed in the literature for describing a time-varying $\Lambda(t)$ vacuum \cite{Overduin1998}. Based on dimensional arguments, Carvalho {\it et al.} \cite{Carvalho1992} shown that a natural dependence is $\Lambda \propto H^{2}$. Later on, Waga added the bare cosmological constant to this $H^{2}$ dependence studying some physical consequences of law $\Lambda (H) = \Lambda_b + \beta H^{2}$, where $\beta$ is a dimensionless free parameter \cite{Waga1993}.  Finally, such a functional dependence was derived within a renormalization group approach based on QFT in curved spacetime by Sol\`a and Shapiro \cite{Shapiro2002}.

In what follows, in order to discuss analytically some subtleties related to the decaying vacuum contribution for the stochastic background of GWs, it will be assumed here that the phenomenological $\Lambda(H)$-term is the one adopted by the above authors \cite{Carvalho1992, Waga1993, Shapiro2002} which is also a particular case of the class studied by the authors of Refs. \cite{Lima2013, Perico2013}:

	\begin{equation} \label{eq7}
	\Lambda(H) =  \Lambda_b + 3\beta H^{2}\,,
	\end{equation}
where $H={\dot a}/a$ is the Hubble parameter, and the factor 3 was added for mathematical convenience. The value of the parameter $\beta$ is not arbitrary, studies suggest that is positive and around $\sim10^{-3}$ \cite{Gomez-Valent:2014rxa}. Assuming the $\omega$-law EoS

	\begin{equation}\label{eq8}
	p=\omega \rho,
	\end{equation}
where $\omega$ is a different constant for each era, and combining Eqs. (\ref{eq5})-(\ref{eq8}), we find that the scale factor is driven by the following differential equation \cite{CL2012}:

	\begin{equation} \label{eq9}
	a\ddot{a}+ \Delta\,\dot{a}^2 - \frac{(1 + \omega)}{2}\Lambda_{b} a^{2}= 0\,,
	\end{equation}
where we have introduced the convenient short notation

	\begin{equation}\label{Delta}
	\Delta=\frac{3(1+\omega)(1-\beta)-2}{2}\,.
	\end{equation}
The case $\beta=0$ reduces to the standard $\Lambda$CDM equation for different eras. In the absence of $\Lambda_b$, the  free $\beta$ parameter was first constrained long back ago based on big-bang nucleosynthesis studies to be $\beta < 0.16$ \cite{Sakhar,Lima1998}. {Actually, it is even smaller when the constant vacuum term is also considered in the same decaying law ($\beta \sim 10^{-3}$) \cite{Sola2016}}.

In terms of the conformal time ($\eta$), the scale factor equation (\ref{eq9}) can be rewritten as (see \cite{Maia1996} for $\beta=0$):

	\begin{eqnarray}\label{a1}
	\frac{a''}{a}+(\Delta-1)\frac{{a'}^2}{a^{2}} - \frac{(1 + \omega)}{2}\Lambda_{b} a^{2}= 0\,.
	\end{eqnarray}
where primes denote derivatives with respect to $\eta$. The last term in the above equation is negligible at early times. Actually this happens for redshifts of the order of few. In this case, the general solution for the scale factor can be written as:

	\begin{equation}\label{scale factor}
	a(\eta)= c_1(\Delta \eta-c_2)^{1/\Delta},
	\end{equation}
where $c_1$ and $c_2$ are integration constants. As usual, the solution for each era is specified by the corresponding value of $\omega$, namely: inflation ($\omega_{\text{inf}}=-1$), radiation-vacuum ($\omega_{\text{rad}}=1/3$) and matter-vacuum ($\omega_{\text{mat}}=0$). For each era we have also assumed that the vacuum decays only on the dominant component. Note also that the early exponential inflation is not modified by the decaying vacuum since $\omega = -1$ implies $\Delta=-1$ so that ${\ddot a}> 0$ regardless of the values of $\beta$ (see Eqs. (\ref{eq9}) and (\ref{Delta})).

In order to find the integration constants for each era we must use the continuity junction conditions for the transition times between each era, $a_n(\eta_i)=a_{n+1}(\eta_i)$ and $a'_n(\eta_i)=a'_{n+1}(\eta_i)$. With this we have:

	\begin{eqnarray}\label{scale factor lambda}
	a(\eta) = \left \{ \begin{matrix}
	-l_i\eta^{-1} \quad \eta<0, \quad \eta\leq\eta_1 \\
    \\
    l_i a_{0r} (\Delta_{\text{rad}}\eta-\eta_{\text{rad}})^{1/\Delta_{\text{rad}}}, \quad \eta_1\leq\eta\leq\eta_{\text{eq}}\\
    \\
    l_i a_{0m} (\Delta_{\text{mat}}\eta-\eta_m)^{1/\Delta_{\text{mat}}}, \quad \eta\geq\eta_{\text{eq}}
    \end{matrix}\right.
	\end{eqnarray}
where $l_i$ is a constant and the parameter $\Delta_{\alpha}$ corresponds to the value $\omega_{\alpha}$. The transition time between inflation and radiation era is $\eta_1$, and between radiation and matter is $\eta_{\text{eq}}$. The values of the integration constants are

	\begin{eqnarray}
	\eta_{\text{rad}} &=& (\Delta_{\text{rad}}+1)\eta_1, \nonumber\\
	a_{0r} &=& (-\eta_1)^{-(1+1/\Delta_{\text{rad}})}, \nonumber\\
	\eta_{\text{mat}} &=& (\Delta_{\text{mat}}-\Delta_{\text{rad}})\eta_{\text{eq}}+\eta_{\text{rad}}, \nonumber\\
	a_{0m} &=& a_{\text{0r}}\frac{(\Delta_{\text{rad}}\eta_{\text{eq}}- \eta_{\text{rad}})^{1/\Delta_{\text{rad}}}}{(\Delta_{\text{mat}}\eta	_{\text{eq}}-\eta_{\text{mat}})^{1/\Delta_{\text{mat}}}},
	\end{eqnarray}
Now we have a complete solution for the scale factor which depends on the parameter $\beta$. In the special case of no decaying vacuum ($\beta=0$), the quantity $\Delta$ as given by (\ref{Delta}) reduces to the standard definition and all the expressions appearing in Ref. \cite{Grishchuk1993} are recovered. For future calculations it is important to estimate values of the transition times $\eta_1$ and $\eta_{\text{eq}}$. In order to do that, let us compare the scale factors of different cosmological eras. First we adopt the convention $a(\eta_0)\equiv 1$ and taking the values of the ratios presented in the Grishchuk work \cite{Grishchuk1993} $a(\eta_0)/a(\eta_1)\simeq 10^{21}$ and $a(\eta_0)/a(\eta_{\text{eq}})\simeq 10^{4}$. This approximated ratios are valid for high redshifts. Using the solution for the scale factor (\ref{scale factor lambda}) and solving the equations system we obtain that $\eta_1 \simeq -10^{-17}$ and $\eta_{\text{eq}} \simeq 3\times10^{-3}$. Although representing a crude approximation since the value of $\eta_{\text{eq}}$ must depend on the $\beta$ parameter, throughout this paper we adopt this values.

\begin{figure}[tbp]
    \centering
    	\includegraphics[scale=0.35]{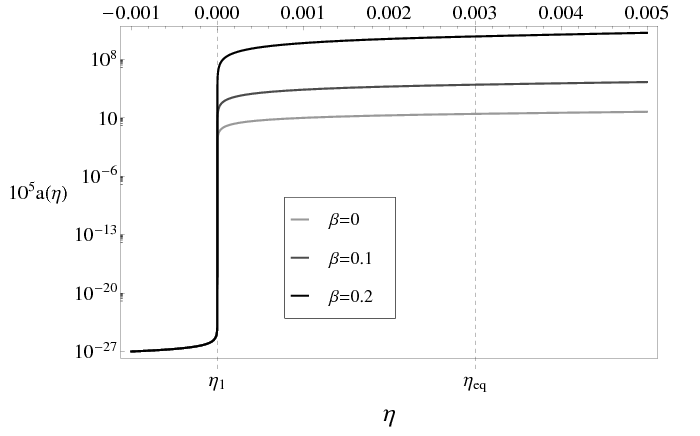}
	\caption{Evolution of the scale factor for different regimes and some selected values of $\beta$ as indicated in the figure.
The  expansion rate is faster for higher values of $\beta$.}
	\label{fig1:scalefactor}
\end{figure}

In the Figure \ref{fig1:scalefactor} we show schematically the behavior of the scale factor for some selected values of $\beta$. As expected, for a given value of $\omega$, the expansion grows faster for higher values of $\beta$. This happens because the vacuum component contributes with a negative pressure.

\section{Cosmological Tensor Perturbations}

In the conformal time, a classical tensor metric perturbation in the FRW flat geometry given by Eq.(\ref{eq1}) can be written as:

	\begin{equation}
	ds^2=a^2(\eta)[-d\eta^2 + (\delta_{ij}+h_{ij})dx^idx^j],
	\end{equation}
where the perturbation $h_{ij}$ is small, $|h_{ij}|\ll 1$, and satisfy the well known \cite{Maggiore2008} symmetry (transverse-traceless) and gauge constraints, namely: $h_{0\mu}=0,\, h^i_i=0, \,\nabla^jh_{ij}=0$.

In the context of GR, the evolution of the tensor perturbations is given by the standard form:

	\begin{equation}
	{h^{j}_i}''+2\frac{a'}{a}{h^{j}_i}'-\nabla^2h^{j}_i=0. \label{heq}
	\end{equation}
This happens because both components in the mixture (matter plus vacuum) have the perfect fluid isotropic form \cite{Weinberg2008,Mukhanov-book}. The general solution can be expressed in terms of Fourier expansion:

	\begin{eqnarray}\label{hij}
	h_{ij}(\eta,\textbf{x}) = \frac{\sqrt{16\pi G}}{(2\pi)^{3/2}}\int d^3\textbf{n}\displaystyle\sum_{r=+,\times} \overset{r}{ \epsilon}_{ij}	(\textbf{n})\left[\overset{r}{h_n}(\eta)e^{i\textbf{n}\cdot\textbf{x}}\,\overset{r}{c}_{\textbf{n}} + \overset{r}{h_n^*}(\eta)e^{-i\textbf{n}	\cdot\textbf{x}}\,\overset{r}{c_{\textbf{n}}}^{\dag}\right],
 \end{eqnarray}
where $\overset{r}{h}_n(\eta)$ are the mode functions, $\textbf{n}$ is the comoving wave vector, $\overset{r}{c_{\textbf{n}}}$ and $\overset{r}{c_{\textbf{n}}}^{\dag}$ are complex numbers. The polarization tensor, $\overset{r}{\epsilon}_{ij}(\textbf{n})$, is symmetric ($\overset{r}{\epsilon}_{ij}(\textbf{n})=\overset{r}{\epsilon}_{ji}(\textbf{n})$), traceless ($\overset{r}{\epsilon}_{ii}(\textbf{n})=0$), and transverse ($n_i \overset{r}{\epsilon}_{ij}(\textbf{n})=0$). We also choose a circular-polarization basis in which $\overset{r}{\epsilon}_{ij}(\textbf{n})=(\overset{r}{\epsilon}_{ij}(\textbf{-n}))^*$, and normalize the basis $\sum_{i,j}\overset{r}{\epsilon}_{ij}(\textbf{n})(\overset{s}{\epsilon}_{ij}(\textbf{n}))^*=2\delta^{rs}$.

The comoving wave number $n=|\textbf{n}|$ is related with the physical wave number $k$ by

	\begin{equation}
	n=|\textbf{n}|=\frac{2 \pi a(\eta)}{\lambda}=k \,a(\eta).
	\end{equation}
Now, by inserting the solution (\ref{hij}) in (\ref{heq}) it is readily seen that the temporal part decouples thereby  giving the evolution equation for the conformal time modes:

	\begin{equation}
	\overset{r}h_n(\eta)''+2\frac{a'}{a}\overset{r}h_n(\eta)'+n^2\overset{r}h_n(\eta)=0.
	\end{equation}
Using the auxiliary function $\overset{r}\mu(\eta,n)=\overset{r}h_n(\eta)a(\eta)$ the above equation assumes the first obtained by Grishchuck \cite{Grishchuk1993} which is independently satisfied for each polarization $r=+,\times$:

	\begin{equation}\label{mu}
	\overset{r}\mu''+\left(n^2-\frac{a''}{a}\right)\overset{r}\mu=0,
	\end{equation}
Therefore, by assuming that the vacuum component is smooth we see that the standard wave equation is not modified. Given the solutions for the scale factor $a(\eta)$ in different eras, we can solve (\ref{mu}) for each mode $n$. It represents an harmonic oscillator with variable frequency determined by the evolution of the Universe and describes different behaviors for the high and low frequency regimes (with and without a vacuum component). Once the solutions for $\mu(\eta)$ for the different cosmic eras has been calculated, it is immediate to obtain the associated  physical quantities like the wave amplitude, energy density and power spectrum.	

\begin{figure}[tbp]
    \centering
    	\includegraphics[scale=0.35]{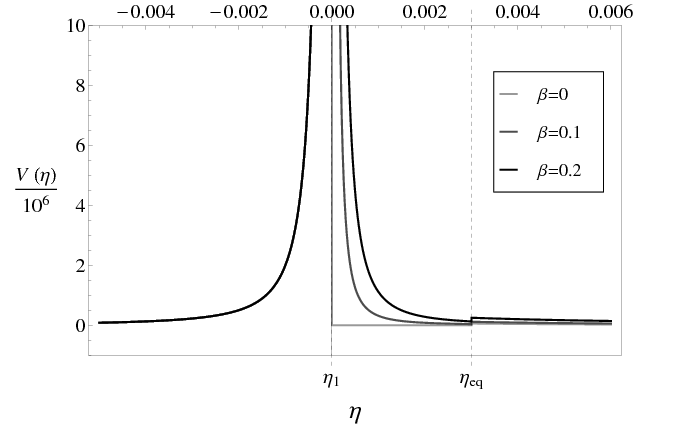}
	\caption{Potential, $V(\eta)=a''/a$ for some values of $\beta$. Note that there is no adiabatic amplification for $\beta =0$ and $\omega=	1/3$ (standard radiation era) since in this case $a''/a \equiv 0$ (see also Eqs. (\ref{Delta})-(\ref{a1})).}
	\label{fig2:potential}
\end{figure}
	
An important quantity driving the behavior of the primordial GWs is the ``potential'' $V(\eta) = a''/a$ appearing in equation (\ref{mu}) (it should be recalled that the name ``potential" come from the mathematical analogue with the stationary Schr\"{o}dinger equation). The relation between the potential and the wave-number determines the behavior of the limit solutions for $\mu(\eta)$.

For the times when $n^2 \gg |V|$ holds, the solution of (\ref{mu}) is oscillatory, $\mu \propto e^{\pm in\eta}$, so that the high-frequency waves are diluted by the cosmic expansion $h=e^{\pm in\eta}/a$. In the opposite limit, $n^2 \ll |V|$, we have $\mu \propto a$, and, consequently, the low-frequency waves obey $h=constant$. The effect of the potential is to avoid the damping of the waves due the universe expansion. The net effect is that the perturbations are relatively enhanced, a phenomenon commonly referred to as \textit{adiabatic amplification} \cite{2Grishchuk1993}. Note that in the limit case ($\beta =0$, $\omega=1/3$), that is, in the standard radiation phase, we see that $\Delta=1$ and, therefore $a'' \equiv 0$. It thus follows that in the radiation era ($\eta_1 < \eta < \eta_{\text{eq}}$), the potential vanishes identically ($V(\eta)\equiv 0$). Physically, this means that there is no adiabatic amplification of GWs during the standard radiation phase.

In Figure \ref{fig2:potential} we show the behavior of the potential for the different eras.  As it will be discussed next, for $\omega=1/3$ and $\beta \neq 0$ GWs are produced so that low frequency modes can be slightly amplified even during the radiation phase. Let us now discuss the solutions of the wave equation for the different eras.

\section{Gravitational Wave solutions for different eras}

\subsection{Inflation Era}

In the particular case of $\omega\equiv\omega_{\text{inf}}=-1$ we have an exponential inflation with a potential, $a''/a=-2/\eta^2$, regardless of the values assumed by the $\beta$ parameter. This case has already been studied in \cite{3Grishchuk1993}. The scale factor $a_{\text{inf}}(\eta) = -l_i/\eta$ gives a positive constant Hubble parameter $H_I = l_i^{-1}$. In the inflation era the GW equation (\ref{mu}) can be written as:

	\begin{equation}\label{eq_mu_inf}
	\mu_{\text{inf}}''+\left(n^2-\frac{2}{\eta}\right)\mu_{\text{inf}}=0,
	\end{equation}
where for simplicity, we have suppressed the polarization index $r$. The general solution of the above equation can be expressed in terms of Bessel's functions

	\begin{equation}
	\mu_{\text{inf}}(n,\eta) = \sqrt{\eta}\,[A_i J_{-3/2}(n\eta)+B_i J_{3/2}(n\eta)].
	\end{equation}
We have to specify some conditions to calculate the integration constants $A_i$ and $B_i$. In the inflation era, the limit for high frequencies must reach the so-called \emph{adiabatic vacuum} $\lim_{n\rightarrow \infty}\mu \propto e^{-in\eta}$ \cite{Birrel-book}. Using this condition and also in its first derivative the constants reduce to $A_i=i\sqrt{\pi/2}$ and $B_i=-\sqrt{\pi/2}$. After doing some algebra we have the normalized ($\mu_{\text{inf}}\,\mu_{\text{inf}}'^*-\mu_{\text{inf}}^*\,\mu_{\text{inf}}'=i$) solution

	\begin{equation}\label{mu inf}
	\mu_{\text{inf}}(n,\eta)=\frac{1}{\sqrt{2n}}\left(1-\frac{i}{n \eta}\right) e^{-i n \eta}.
	\end{equation}
Knowing the full expressions for $\mu$ and $a$ is easy to calculate the power $\mathcal{P}$ and energy spectrum $\Omega_{\text{gw}}$. For details see the Appendix B.
	\begin{eqnarray}
	\mathcal{P}_{\text{inf}}(n,\eta) &=& \frac{16 G}{\pi l_i^2}(1+n^2\eta^2), \nonumber\\
	\Omega_{\text{gw}}^{(\text{inf})}(n,\eta) &=& \frac{8G}{3\pi l_i^2} n^4\eta^4\left(2+\frac{1}{n^2\eta^2}\right).
	\end{eqnarray}
These are standard results already studied, for details see \cite{Mukhanov-book}. In particular for long wavelengths $\lambda \gg l_i = H^{-1}_I$ the power spectrum is flat and proportional to $H_I$, $\mathcal{P}_{\text{inf}} = \frac{16 G}{\pi}H^2_i$. For our purposes the calculation of $\mu_{\text{inf}}$ is important to obtain a complete solution for $\mu_{\text{rad}}$ in the radiation era that will be shown in the next section.

\subsection{Radiation Era}

In the radiation era we have $\omega\equiv\omega_{\text{rad}}=1/3$  and  $\Delta_{\text{rad}} = 1-2\beta$. From the second solution given by Eq. (\ref{scale factor lambda}), it is readily seen that the Eq. (\ref{mu}) now takes the form:

	\begin{equation}
	\mu_{\text{rad}}'' + \left(n^2 - \frac{1-\Delta_{\text{rad}}}{[\Delta_{\text{rad}}(\eta-\eta_1)-\eta_1]^2} \right)\mu_{\text{rad}} = 0\,.
	\end{equation}
	
At this point, we have to stress the first important result of the present paper: The potential for this case is $V=-2\beta (a'/a)^2 \neq 0$. In the particular case of $\beta=0$ (no decaying vacuum) we obtain the well-known result $V=0$ of no GW amplification in the radiation era. In the decaying vacuum models this does not hold anymore, the $V\neq0$ condition implies that always the radiation will contribute to the primordial GW spectrum today.
   	
The general solution of the last equation is:

	\begin{eqnarray}
\mu_{\text{rad}}(n,\eta) &=& \sqrt{\Delta_{\text{rad}}\eta-\eta_1(\Delta_{\text{rad}}+1)} \\
	&\times& \left[ A_{r} J_{\alpha_{r}}\left(\frac{n}{\Delta_{\text{rad}}}(\Delta_{\text{rad}}\eta-\eta_1(\Delta_{\text{rad}}+1))\right) + B_{r} J_{\alpha_{r}} \left(\frac{n}{\Delta_{\text{rad}}}(\Delta_{\text{rad}}\eta-\eta_1(\Delta_{\text{rad}}+1))\right.	\right]	\nonumber\,,
	\end{eqnarray}
where $\alpha_{r}=\frac{1}{\Delta_{\text{rad}}}-\frac12$. The continuity junction conditions, $\mu_{\text{inf}}(\eta_1)=\mu_{\text{rad}}(\eta_1)$ and $\mu'_{\text{inf}}(\eta_1)=\mu'_{\text{rad}}(\eta_1)$, must be used  to calculate the integration constants $A_r$ and $B_r$. Making the calculations we have:

	\begin{eqnarray}
	A_{r} &=& \frac{e^{-in\eta_1}\pi\sec(\pi/\Delta_{\text{rad}})}{\Delta_{\text{rad}} (-2n\eta_1)^{3/2}} \left[n\eta_1(i-n\eta_1)J_{1-\alpha	_{r}}\left(\eta*\right)- (2i - n\eta_1(2+in\eta_1) - \Delta_{\text{rad}}(i-k\eta_1)) J_{-\alpha_{r}}\left(\eta*\right)\right],\nonumber\\
	B_{r} &=& \frac{e^{-in\eta_1}\pi\sec(\pi/\Delta_{\text{rad}})}{\Delta_{\text{rad}} (-2n\eta_1)^{3/2}} \left[-in^2\eta_1^2 J_{\alpha_{r}}	\left(\eta*\right)(in\eta - n^2\eta^2_1)J_{\alpha_{r}+1}\left(\eta*\right)\right].\nonumber
	\end{eqnarray}
where $\eta* = -\frac{n\eta_1}{\Delta_{\text{rad}}}$.
Note that the above solution for the radiation era is also normalized since it satisfies $\mu_{\text{rad}}\,\mu_{\text{rad}}'^*-\mu_{\text{rad}}^*\,\mu_{\text{rad}}'=i$. These expressions allow us to obtain  the amplitude of the perturbations.

\begin{figure}[tbp]
    \centering
    	\includegraphics[scale=0.2]{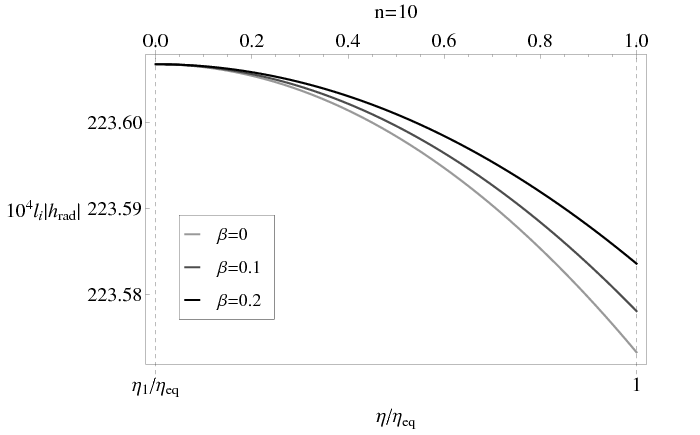}
        \includegraphics[scale=0.2]{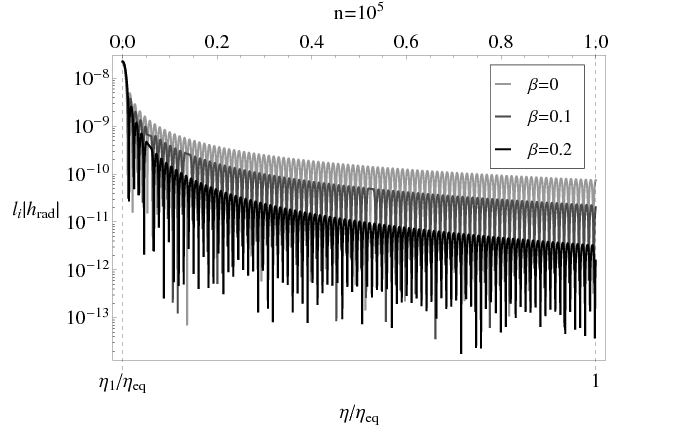}
        \includegraphics[scale=0.2]{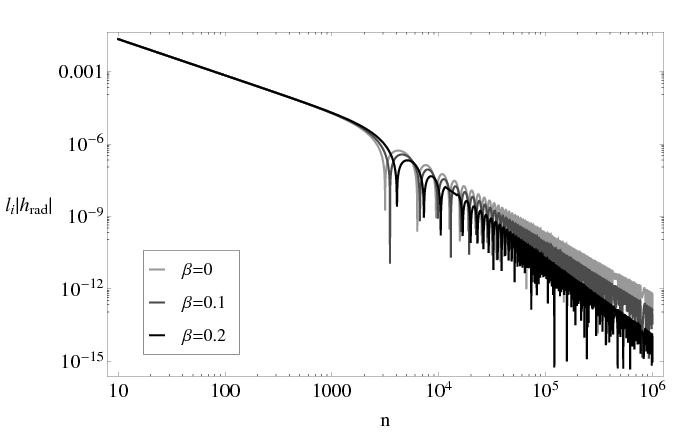}
	\caption{The amplitude of the GW in the radiation era. {\bf{a)}} Modulus of the mode function $|h_{\text{rad}}|$ as a 	function of the conformal time for some selected values of $\beta$ and a fixed comoving low frequency, $n=10$. {\bf{b)}} The same plot of figure {\bf a}) but now for a fixed high frequency, $n=10^{5}$. Note that in the low frequency regime decaying vacuum models amplify the perturbations since the amplitude is higher in comparison to the case $\beta=0$. However, high frequency modes are always damped out regardless of the value of $\beta$ (see main text). {\bf{c)}} Modulus of the mode function $|h_{\text{rad}}|$ as a function of $n$ and the same selected values of $\beta$ at $\eta_{\text{eq}}$.}
	\label{fig3:rad amplitude}
\end{figure}

Figures \ref{fig3:rad amplitude}a and \ref{fig3:rad amplitude}b display the evolution of the amplitude $|h_{\text{rad}}|$ as a function of the conformal time $\eta$ and some selected values of $\beta$. As discussed above, the behavior of low frequency (Figure \ref{fig3:rad amplitude}a) and high frequency (Figure \ref{fig3:rad amplitude}b) modes are quite different since the former are amplified (even during the radiation phase) while the later are damped. Surprisingly, we see that the high frequencies modes are damped even faster than in the standard case ($\beta = 0$).

How can such a result be understood? The basic point here is that in the high frequency limit the term $a''/a$ can be neglected and $h (\eta) = a^{-1}(\eta)\mu(\eta)$. Therefore, since the solution of $\mu$ is an oscillating function, the amplitudes are damped out even more intensively (in comparison to $\beta=0$) since  the scale factor expands faster for higher values of $\beta \neq 0$ (see Figure \ref{fig1:scalefactor}).

Conversely, in the low frequency regime we find exactly the opposite behavior. Indeed, due to the condition $n^2 \ll a''/a$, the solution for low frequencies is $\mu \propto a$ so that the perturbations, $h(\eta)$, remains nearly constant (see Figure 3a). Is exactly in this regime (low frequencies) that the amplification occurs even during the radiation phase. Finally, in figure 3c we show the behavior of $|h_{\text{rad}}|$ as a function of the frequency for a fixed time.

\subsection{Matter Dominated era}

In the matter dominated phase, the EoS is $\omega\equiv\omega_{\text{mat}}=0$ which implies that $\Delta_{\text{mat}} = (1-3\beta)/2$.
In this case, the scale factor reduces to:

\begin{equation}
a_{\text{mat}}(\eta) = l_i\, a_{0m} \left(\frac{1-3\beta}{2}\eta-\eta_{\text{mat}}\right)^{2/(1-3\beta)}
\end{equation}
where $\eta_{\text{mat}} = -\frac{1-\beta}{2}\eta_{\text{eq}}+2(1-\beta)\eta_1$ and $a_{0m} = (-\eta_1)^{-\frac{2(1-\beta)}{1-2\beta}}\left[(1-2\beta)\eta_{\text{eq}}-2(1-\beta)\eta_1\right]^{-\frac{1-\beta}{1-5\beta+6\beta^2}}$. Calculating the potential and substituting in (\ref{mu}) we have

\begin{eqnarray}
\mu_{\text{mat}}'' + \left(n^2 - \frac{2+6\beta}{[(3\beta-1)\eta+2\eta_{\text{mat}}]^2} \right)\mu_{\text{mat}} = 0,
\end{eqnarray}
solving it we obtain the general solution

\begin{eqnarray}
\mu_{\text{mat}} &=& \sqrt{\Delta_{\text{mat}}\eta-\eta_{\text{mat}}} \left[A_{m} J_{\alpha_{m}}\left(\frac{n (\Delta_{\text{mat}}\eta-\eta_{\text{mat}})}{\Delta_{\text{mat}}}\right)+ B_{m} J_{-\alpha_{m}}\left(\frac{n (\Delta_{\text{mat}}\eta-\eta_{\text{mat}})}{\Delta_{\text{mat}}}\right)\right],
\end{eqnarray}
with the index $\alpha_{m}=\frac{1}{\Delta_{\text{mat}}}-\frac12$. Again the constants, $A_{m}$ and $B_{m}$, are obtained by using the continuity conditions at the transition time $\eta_{\text{eq}}$ between the radiation and matter era, $\mu_{\text{rad}}(\eta_{\text{eq}}) = \mu_{\text{mat}}(\eta_{\text{eq}})$ and $\mu_{\text{rad}}'(\eta_{\text{eq}}) = \mu_{\text{mat}}(\eta_{\text{eq}})'$, the full expressions are cumbersome and will not be presented.

\begin{figure}[tbp]
    \centering
    	\includegraphics[scale=0.2]{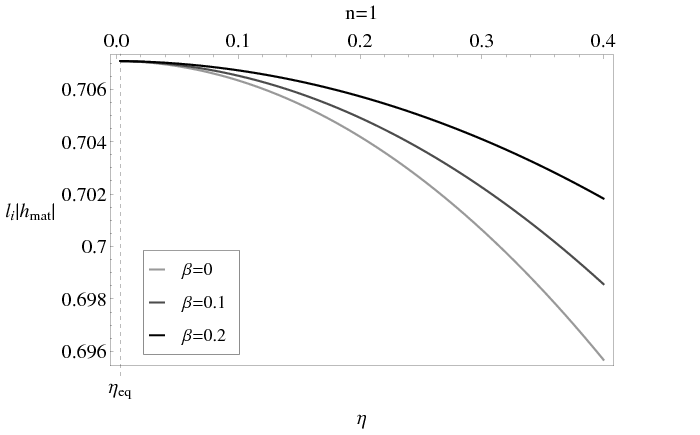}
        \includegraphics[scale=0.2]{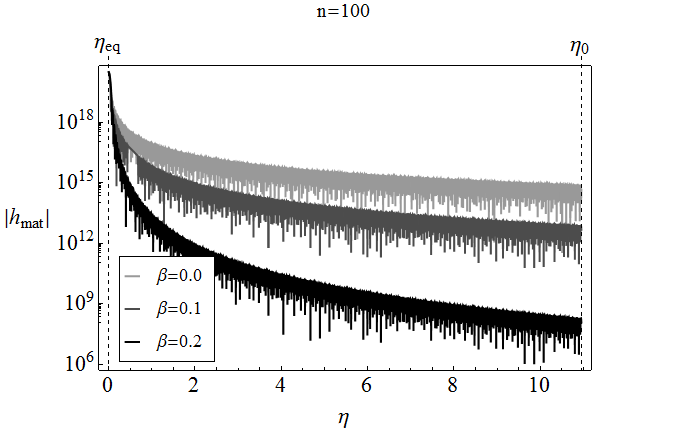}
        \includegraphics[scale=0.2]{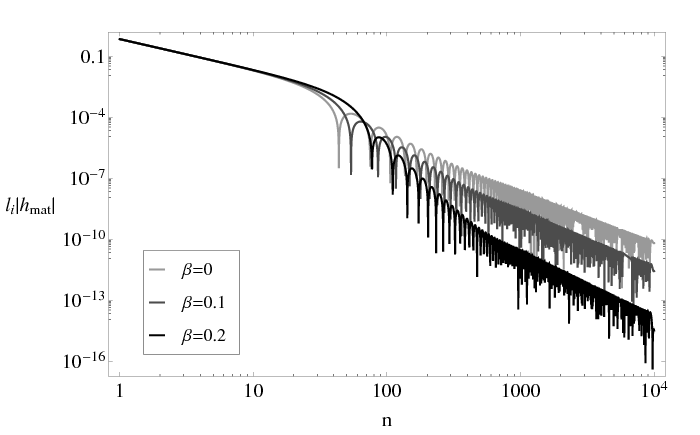}
	\caption{The amplitude of the GW in the matter era. {\bf{a)}} Modulus of the mode function $|h_{\text{mat}}|$ as a 	function of the conformal time for some selected values of $\beta$ and a fixed  comoving low frequency, $n=1$.  {\bf{b)}} Modulus of the mode function $|h_{\text{mat}}|$ for a high frequency, $n=10^{4}$. {\bf{c)}} Modulus of the mode function $|h_{\text{mat}}|$ as a function of $n$ and the same selected values of $\beta$ at $\eta_{\text{0}}$.}
	\label{fig4:mat amplitude}
\end{figure}

In Figure \ref{fig4:mat amplitude} we plot $|h_{\text{mat}}|$. We can see that a behavior like $|h_{\text{rad}}|$ is also obtained. As one may conclude, this happens because of the same reasons already discussed in the preceding section. For low frequencies if we have two perturbations $|h_1(\beta_1)|$ and $|h_2(\beta_2)|$, with  $\beta_1 >\beta_2$, then $|h_1|>|h_2|$. This condition inverts for the high frequency regime (for $\beta_1 >\beta_2$ then $|h_1|<|h_2|$).

\section{Power and energy density spectra}

Let us now discuss the power spectrum and the spectral energy density parameter (per logarithmic wave number interval) which can be written as (see Appendix for details):

\begin{equation}\label{power spectrum}
    \mathcal{P}(n,\eta)  =  \frac{32 G}{\pi} n^3 |h_n(\eta)|^2 \,,
 \end{equation}

and

  \begin{equation}\label{energy spectrum}
    \Omega_{\text{gw}}(n,\eta) = \frac{8\pi G}{3\mathcal{H}^2(\eta)}\frac{n^3}{2 \pi^2}(|h'_n(\eta)|^2+n^2|h_n(\eta)|^2).
    \end{equation}

In comparison to other alternative cosmologies, one advantage of our simple decaying vacuum models is that the above quantities can analytically be calculated for the radiation and matter dominated eras. Actually, this happens because the amplitude $h(n, \eta)=\mu/a$ and the corresponding solutions of $\mu$ were explicitly obtained for each case (see previous section).

\begin{figure}[tbp]
    \centering
    	\includegraphics[scale=0.3]{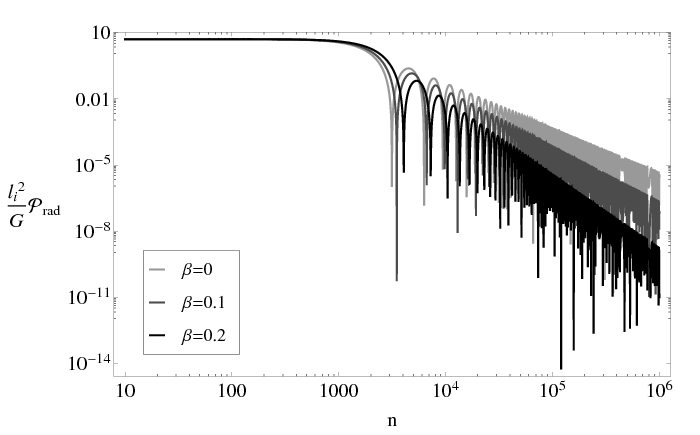}
        \includegraphics[scale=0.3]{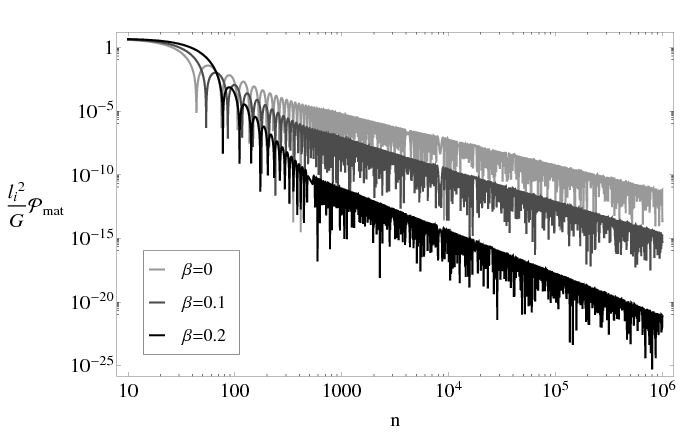}
	\caption{The power spectra of the GWs in the radiation and matter dominated eras at $\eta_{\text{eq}}$ and $\eta_0$ respectively.}
	\label{fig5:power spectra}
\end{figure}

In Figure \ref{fig5:power spectra} we show the plots of the power spectra, $\mathcal{P}_{\text{rad}}$ and $\mathcal{P}_{\text{mat}}$, for radiation and matter dominated eras, respectively. From the first plot we see that $\mathcal{P}_{\text{rad}}$ is almost flat until some transition frequency when begins to decrease.
Similarly to what happens for the amplitudes, the power spectrum in this regime is slightly larger as the parameter $\beta$ increases. As should be expected,  the decaying vacuum contributes to the creation of low-frequency gravitons with the corresponding spectrum remaining essentially flat.

Nevertheless, after some transition frequency (corresponding to $n \sim few\times 10^{3}$), the waves are strongly damped and the associated power spectrum is no longer flat. This means that the decaying vacuum in this regime contributes more to increase the scale factor than to the production of gravitons. $\mathcal{P}_{\text{rad}}$ decreases exponentially and the effect is even more pronounced for larger values of  $\beta$. Note also that at late times, the  $\mathcal{P}_{\text{mat}}$ for the matter-vacuum dominated phase presents the same high and low frequencies general properties of the radiation era. It starts with an almost flat spectrum and also decreases faster as long as the vacuum contribution is relatively larger (higher values of $\beta$).

\begin{figure}[tbp]
    \centering
    	\includegraphics[scale=0.3]{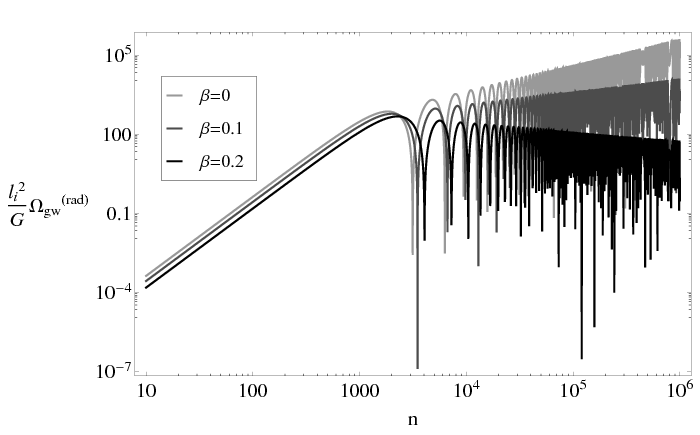}
        \includegraphics[scale=0.3]{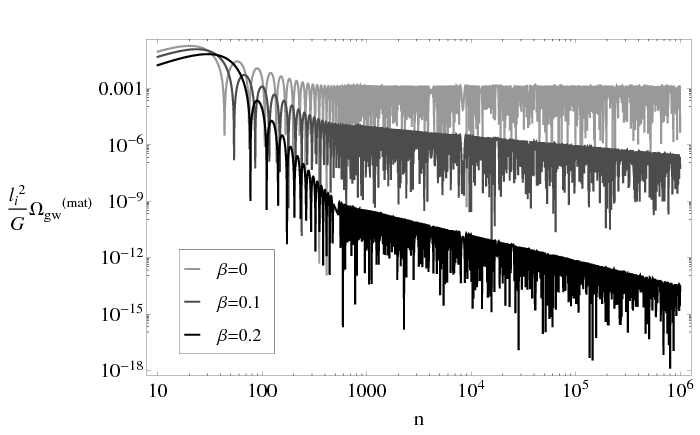}
	\caption{The energy density spectra of the GWs for the radiation and matter dominated eras at $\eta_{\text{eq}}$ and $\eta_0$ respectively.}
	\label{fig6:energy density spectra}
\end{figure}

In Figure \ref{fig6:energy density spectra} we display the energy density parameter as a function of the comoving wave number for the radiation and matter dominated eras.  An interesting feature of the energy density spectrum in the radiation era, $\Omega_{\text{gw}}^{(\text{rad})}$, is that if $\beta_1>\beta_2$ then $\Omega_{\text{gw}}^{(\text{rad})}(\beta_1) <  \Omega_{\text{gw}}^{(\text{rad})}(\beta_2)$ for all frequencies. Note also that $\Omega_{\text{gw}}^{(\text{rad})}$ grows as a power-law being weakly dependent on the value of $\beta$, but a more strong dependence is obtained at the high frequency limit.

The evolution of $\Omega_{\text{gw}}^{(\text{mat})}$ follows a similar trends with some peculiarities at the high frequency limit. As in the radiation case if $\beta_1>\beta_2$ then $\Omega_{\text{gw}}^{(\text{mat})}(\beta_1) < \Omega_{\text{gw}}^{(\text{mat})}(\beta_2)$ for all frequencies. In addition, for low frequencies the spectrum also grows as a power law linear being slightly lower for bigger values of $\beta$. However, in the high frequency regime, the spectrum always decreases but varies differently as a function of the $\beta$ parameter. In particular, for $\beta \sim 0.2$, the fall is very abrupt. Note also that for $\beta=0$ it initially decreases and after remains almost flat for all modes. The basic reason for such a behavior is simple:  when $|h'|^2 \ll |h|^2$ the energy density spectrum is $\Omega_{\text{gw}} \propto n^5|h|^2\mathcal{H}^{-2}$, where $\mathcal{H}=a'/a$ (see Eq. (\ref{energy spectrum})). This means that the contribution of the factor $\mathcal{H}^{-2}$ (which is not present in the power spectrum) largely determines the behavior of $\Omega_{\text{gw}}^{(\text{mat})}$.

\section{Final Comments}

In this paper, by assuming a spatially flat geometry, we have investigated the production of GWs for an interacting mixture of matter and vacuum in the context of GR. The dynamical $\Lambda$-term was described by a phenomenological law: $\Lambda (H) = \Lambda_b + 3\beta H^2$, and a three stage description involving inflation, radiation and matter eras was adopted. It was also assumed that the vacuum decays only on the dominant component.

For each cosmic era, we have determined the general expressions for the scale factor (in the conformal time), as well as the analytical solutions for the GW equation (see Figs. \ref{fig1:scalefactor}, \ref{fig3:rad amplitude} and \ref{fig4:mat amplitude}). We notice that the mode function equations of the primordial GWs were derived and explicitly solved for each era. More interesting, the corresponding power spectra and the energy density parameter for the radiation and matter era were also obtained (see Figs. \ref{fig5:power spectra} and \ref{fig6:energy density spectra}). Obviously, exact solutions are allowed in this framework due to the simplified phenomenological form adopted for the  decaying  $\Lambda(H)$-term.

In the present running vacuum model, the scale factor expands faster as long as the $\beta$ parameter increases thereby affecting considerably the production and evolution of the GW modes for both regimes - low and high frequency limits (see Figs. \ref{fig3:rad amplitude} and \ref{fig4:mat amplitude}). However, the most prominent feature is that the ``potential''  in the radiation era never vanishes (see Fig. \ref{fig2:potential}), even when very small values of the $\beta$ parameter are adopted. As a consequence, unlike models with no  decaying vacuum ($\beta=0$),  the GWs in the radiation era can be adiabatically amplified (in the sense of Grishchuk \cite{2Grishchuk1993}). This is the main result of the paper. It is different from the standard Parker's result \cite{Parker1968, Parker1969} claiming that in relativistic cosmology there is no gravitationally induced quantum production of particles due the cosmic expansion in the the FRW radiation phase (for a similar result outside of GR see \cite{Pereira2010}. It is closely related with the fact that during the radiation phase $a''\neq 0$ in the presence of a decaying vacuum component.

On the other hand, our expressions for the modulus of the GW amplitude $|h|$ shown us that the adiabatic amplification (graviton production), is a low-frequency phenomenon even for this kind of $\Lambda(H)$ cosmologies. This interesting and known feature is also maintained  in the present context even considering that GW are produced in the radiation phase. Here as there, the basic problem is that at the high frequency regime the cosmic expansion dominates, and, therefore, the perturbations are dynamically suppressed in the course of the expansion. This behavior is also reproduced in the power spectrum due $\mathcal{P} \propto n^3 |h|^2$. However, in the case of the energy density parameter spectrum, $\Omega_{\text{gw}} \propto \mathcal{H}^{-2} n^2 \mathcal{P}$, the behavior is different due to the contribution of $\mathcal{H}^{-2}$.

Finally, we stress that the present work was based on a very simple decaying vacuum cosmology. It played the role of a toy model for obtaining analytically some basic information, like the production of GWs during the radiation phase. Its consequences on the B-modes polarization of CMB anisotropies it will discussed in a forthcoming communication. As it appears, the model can be thought as a starting point for the investigation of  more complex and rich decaying vacuum cosmologies, like the one proposed in Refs. \cite{Lima1994, 2Lima1996}, and recently, investigated in a more enlarged framework in Refs. \cite{Lima2013, Perico2013}. Although less analytical regarding  the calculations involving GWs, such models deserve a closer scrutiny since they furnish a complete cosmological history.

\appendix

\section{Quantized tensor perturbations}

The generating mechanism of the primordial GWs is believed to have a quantum mechanical origin. The basic idea is that quantum fluctuations of the vacuum state in the early universe were stretched to macroscopic scales due the cosmic inflation thereby originating the present observable primordial GW spectrum. The standard quantization procedure is based on a semi-classical approach where the perturbations are quantized but the gravitational background evolves classically (for details see the review of Giovannini \cite{Giovannini2010}).

When the perturbations are quantized on a classical background, the functions $\overset{r}{c_{\textbf{n}}}$ e $\overset{r}{c_{\textbf{n}}}^{\dag}$ in (\ref{hij}) are promoted to be quantum creation and annihilation operators which satisfy equal time commutation relations ($\hbar=1$)

	\begin{equation}\label{commutator}
	[\overset{r'}{c_{\textbf{n}}},\overset{r}{c_{\textbf{m}}}^{\dag}]=\delta_{r'r}\delta^3(\textbf{n}-\textbf{m}), \quad \overset{r}{c}_{			\textbf{n} }|0\rangle_{\textbf{n}}=0.
	\end{equation}

The vacuum state is defined for a given time $\eta$ and mode $\bf{n}$, $|0(\eta)\rangle_{\textbf{n}}$. In general does not exist an unique vacuum state. For a fixed time $\eta_i$ the vacuum state is $\overset{r}{c}_{\textbf{n}}(\eta_i) |0_i\rangle_{\textbf{n}}=0$ (notation $|0(\eta_i)\rangle_{\textbf{n}} = |0_i\rangle_{\textbf{n}}$ ), for another time $\eta_f$ the operator acts differently to the same state which result in general is not null, $\overset{r}{c}_{\textbf{n}}(\eta_f)|0_i\rangle_{\textbf{n}}\neq0$. For an expanding background, the vacuum state at time $\eta_i$ is different from the vacuum state at $\eta_f$. This so-called vacuum state ambiguity has some interesting consequences, the most prominent of them being  the gravitationally-induced particle production (`Grishchuk particles'). To be more precise, by assuming that there is no particles at time  $\eta_i$, that is, $N_i |0_i\rangle=0$, where $N_i$ is the number operator acting in the vacuum state  [$N_i \equiv c^{\dag}_i c_i$ (the polarization and wave number index were suppressed for simplicity)]. Later on, at a time $\eta_f$ we have $N_f |0_f\rangle=0$, but the vacuum state ambiguity gives $N_f |0_i\rangle\neq 0$. Consequently the vacuum state $|0_i\rangle$ contains ``$f$" particles and vice versa. This is the physical foundation for the creation of `Grishchuk gravitons' induced by the cosmic expansion.

\section{Power and energy density spectra}

With the perturbations already quantized, the important physical observables are readily calculated. The (dimensionless) power spectrum, that is, the quadratic mean value of the amplitude of the perturbations, can be defined as:

    \begin{equation}
    \mathcal{P}(n,\eta) \equiv \frac{d \langle0|h_{ij}(\eta,\textbf{x})h^{ij}(\eta,\textbf{x})|0\rangle}{d \ln n}\,,
    \end{equation}
Now, by  inserting (\ref{commutator}) into (\ref{hij}) one finds

    \begin{eqnarray}
    \langle0|h_{ij}(\eta,\textbf{x})h^{ij}(\eta,\textbf{x})|0\rangle = \frac{32 G}{\pi} \int_0^{\infty} n^3 |h_n(\eta)|^2 d\ln n \,.
    \end{eqnarray}
Note that the equality holds because the modulus of the modes for different polarizations are equal, $|\overset{+}{h}|^2 = |\overset{\times}{h}|^2\equiv |h|^2$. Finally we have:

    \begin{eqnarray}\label{power spectrum}
    \mathcal{P}(n,\eta)  =  \frac{32 G}{\pi} n^3 |h_n(\eta)|^2.
    \end{eqnarray}
Another important quantity is the energy spectrum defined in the following manner:

    \begin{equation}\label{ES}
    \Omega_{\text{gw}}(n,\eta) \equiv \frac{1}{\rho_{\text{crit}}}\frac{d \langle0|\rho_{\text{gw}}(\eta) |0\rangle}{d \ln n}\,,
    \end{equation}
which represents the GW energy density ($\rho_{\text{gw}}$) per logarithmic wave number interval, in units of the critical density $\rho_{\text{crit}}(\eta) = 3 H^2(\eta)/8 \pi G$. The GW density is

    \begin{equation}
    \rho_{\text{gw}} = T^0_0 = \frac{1}{64 \pi G}\frac{(h'_{ij})^2+(\nabla h_{ij})^2}{a^2}\,,
    \end{equation}
whose vacuum expectation value in the vacuum state reads:

    \begin{equation}
    \langle0|\rho_{\text{gw}}|0\rangle = \int^{\infty}_0 \frac{n^3}{2 \pi^2}\frac{|h'_n(\eta)|^2+n^2|h_n(\eta)|^2}{a^2}\frac{dn}{n}\,,
    \end{equation}
while the energy spectrum (\ref{ES}) becomes

    \begin{equation}\label{energy spectrum}
    \Omega_{\text{gw}}(n,\eta) = \frac{8\pi G}{3\mathcal{H}^2(\eta)}\frac{n^3}{2 \pi^2}(|h'_n(\eta)|^2+n^2|h_n(\eta)|^2).
    \end{equation}
Both quantities are the most important primordial GW observables.

\acknowledgments

The authors are  grateful to J. C. N. de Araujo and M. E. S. Alves for helpful discussions. DT and JASL acknowledge CAPES and CNPq (Brazilian Research Foundations) for partial support.

\end{document}